\documentclass[11pt]{article}
\usepackage{amsmath}
\usepackage{cite} 
\usepackage{indentfirst}
\usepackage{latexsym}

\long\def\ca#1\cb{} 

\newcommand{\AND}{\mbox{\small AND}}

\newcommand{\becs}{\begin{cases}}
\newcommand{\bem}{\begin{matrix}}

\newcommand{\encs}{\end{cases}}
\newcommand{\enm}{\end{matrix}}

\newcommand{\lra}{\leftrightarrow }

\newcommand{\ra}{\rightarrow }
\newcommand{\Ra}{\Rightarrow }

\def\outl#1{\par{\medskip\noindent\hspace*{.2cm}\bf
      \mathversion{bold}#1\mathversion{normal}\smallskip} }
 \def\xa{} \def\xb{}  

 \def\outl#1{}  \def\xa{} \def\xb{}  

\ca
 \def\outl#1{\par{\medskip\noindent\hspace*{.5cm}\bf
      \mathversion{bold}#1\mathversion{normal}\smallskip} }
 \long\def\xa#1\xb{}
 
\cb

\begin{document} 

\title{Comment on ``Loophole-free Bell inequality''}

\author{Robert B. Griffiths
\thanks{Electronic mail: rgrif@cmu.edu}\\ 
Department of Physics,
Carnegie-Mellon University,\\
Pittsburgh, PA 15213, USA}
\date{Version of 4 December 2015}
\maketitle  
\ca
\centerline{Robert B. Griffiths}
\centerline{Physics Department}
\centerline{Carnegie-Mellon University}
\vspace{.2cm}
\cb

\xb

\xa
\begin{abstract}
  A recent experiment yielding results in agreement with quantum theory and
  violating Bell inequalities was interpreted [Nature 526 (29 Octobert 2015)
  p.~682 and p.~649] as ruling out any local realistic theory of nature. But
  quantum theory itself is both local and realistic when properly interpreted
  using a quantum Hilbert space rather than the classical hidden variables used
  to derive Bell inequalities. There is no spooky action at a distance in the
  real world we live in if it is governed by the laws of quantum mechanics.
\end{abstract} 
\xb



\begin{center}
\textbf{Comment}
\end{center}

\xb
\outl{Delft article claim: Bell proved 'locality + realism' contrary to  QT}
\xa

\xb \outl{Hilbert space QM (HSQM) is local \& realistic. Bell in fact
  showed: (HV + local) $\neq$ QT} \xa

One can admire the technical skill that went into planning and carrying out
the experiment reported in \cite{Hnao15} without necessarily agreeing
with the some of the conclusions drawn by the authors or found in the
accompanying commentary \cite{Wsmn15}. At the beginning of the
abstract of \cite{Hnao15} one finds the assertion: 
\begin{quote}
  More than 50 years ago, John Bell proved that no theory of nature that
  obeys locality and realism can reproduce all the predictions of quantum
  theory: in any local-realist theory, the correlations between outcomes of
  measurements on distant particles satisfy an inequality that can be violated
  if the particles are entangled.
\end{quote}
On the contrary, there is a theory of nature that is both local and realistic,
and reproduces all the predictions of quantum theory. It is known as
\emph{quantum mechanics} or, to be more precise, let us call it
\emph{Hilbert-space quantum mechanics}. What John Bell actually proved was that
a theory based upon \emph{hidden variables} rather than the quantum Hilbert
space, and satisfying an additional assumption of locality, makes predictions
(Bell inequalities) that disagree with quantum theory, and with a series of
experiments of increasing precision, accuracy and sophistication, of which
those reported in \cite{Hnao15} are among the most recent. These experimental
results rule out hidden variables, while being perfectly compatible with the
locality and realism of Hilbert space quantum mechanics.

\xb
\outl{Noncommutation separates QM, CM.  Fine(1982): HV $\lra$ probabilities
rejected by  QM }
\xa

In his first course in quantum physics the student learns that $PX$ is not the
same thing as $XP$: quantum operators associated with physical quantities, such
as momentum and position, in general \emph{do not commute}. Such noncommutation
is the most fundamental way in which quantum theory differs from classical
mechanics, and the connection of this with Bell (and CHSH) inequalities was
pointed out by Fine in 1982 \cite{Fne82b}:

\begin{quote}
  \dots I believe that [the material presented earlier in the article]
  shows what hidden variables and the Bell inequalities are all about; namely,
  imposing requirements to make well defined precisely those probability
  distributions for noncommuting observables whose rejection is the very
  essence of quantum mechanics.
\end{quote}
In other words, Bell inequalities based on hidden variables employ probabilities
in a manner inconsistent with the principles of quantum mechanics. Let us 
look at this in more detail.

\xb
\outl{Spin half: $S_z=\pm1/2\,\lra$ sample space to which probabilities can be
  assigned}
\xa

A spin-half particle is the simplest example of a system whose correct
description requires the use of a Hilbert space,%
\footnote{The term ``Hilbert space'' was first used in the study of
  infinite-dimensional spaces of functions, but is nowadays also
  employed for a finite-dimensional space.} %
in this case a two-dimensional complex vector space with an inner product. The
$x$ and $z$ components of angular momentum of such a particle, $S_x$ and $S_z$,
are operators (matrices) whose eigenvalues are $+1/2$ and $-1/2$ in units of
$\hbar$. Probabilities can be assigned to the two possible values of $S_z$
because they are mutually exclusive and exhaust all possibilities: $S_z$ takes
no values apart from $+1/2$ and $-1/2$. In standard (Kolmogorov) probability
theory an exhaustive set of mutually-exclusive possibilities is called a
\emph{sample space}, and probabilities are assigned to elements of the
sample space, or sets of elements from the sample space. From the physics point
of view it was the Stern-Gerlach experiment that first identified this
sample space.

\xb
\outl{$S_x=\pm 1/2$ also a sample space; cannot combine $S_x$, $S_z$ values}
\xa

Similarly, $S_x$ can take on only the values $+1/2$ and $-1/2$, and these again
constitute a sample space. Were one concerned with classical and not quantum
physics, it would be sensible to talk about a joint probability distribution
based on a sample space of the four mutually exclusive possibilities in which
both $S_z$ and $S_x$ have well-defined values: $S_z=+1/2\ \AND\ S_x=+1/2$,\
$S_z=+1/2\ \AND\ S_x=-1/2$, etc. This is the assumption made, explicitly or
implicitly, in various hidden-variables models. But in Hilbert-space quantum
mechanics as developed by von Neumann (who, incidentally, invented the term
`Hilbert space') there is no such thing as $S_z=+1/2\ \AND\ S_x=+1/2$. The
reason is that quantum properties are associated with \emph{subspaces} of the
Hilbert space; see Sec.~III.5 of \cite{vNmn55}. Within the two-dimensional
Hilbert space of a spin-half particle there is a one-dimensional subspace
associated with $S_z=+1/2$, another with $S_z=-1/2$, another with $S_x=+1/2$,
and so forth; a mathematically-distinct subspace is associated with every
direction in physical space. But this Hilbert space contains no one-dimensional
subspace that can be associated with $S_z=+1/2\ \AND\ S_x=+1/2$, or with any of
the three other possibilities needed if one is to assign a joint probability
distribution to $S_x$ and $S_z$; this reflects the fact that they are
\emph{incompatible} observables, the operators do not commute.%
\footnote{In the quantum logic of Birkhoff and von Neumann \cite{BrvN36},
  $S_z=+1/2\ \AND\ S_x=+1/2$ is assigned the zero-dimensional subspace
  consisting of the zero vector, a property which is always false. Alas,
  quantum logic has not turned out to provide an approach to quantum mechanics
  useful for giving it a physical interpretation. The fundamental difficulty as
  seen from a physicist's perspective is discussed for the case of a spin-half
  particle in Sec.~4.6 of \cite{Grff02c}.} %

\xb \outl{Students taught: cannot measure both $S_x,S_z$. Should be taught:
  combination does not exist} \xa

\xb
\outl{CHSH derivation assumes $S_x,S_z$ have joint prob distribution}
\xa

The preceding remarks agree with what students are taught in an introductory
quantum course: it is impossible to simultaneously \emph{measure} $S_x$ and
$S_z$ for a spin-half particle. Alas, they are rarely told the reason behind
this: there is no property represented by a subspace in the quantum Hilbert
space that corresponds to $S_x$ and $S_z$ simultaneously having specific
values. And even skilled experimentalists cannot measure what does not exist.
However the derivation of the CHSH version of a Bell inequality, which is
Eq.~(1) in \cite{Hnao15}, has as one of its assumptions that $S_x$, $S_z$, and
other components of spin can be replaced by classical, which is to say
commuting, quantities, which have a joint probability distribution, directly
contrary, as Fine pointed out, to the principles of quantum mechanics. That an
inequality based on such assumptions is violated by experimental results is not
surprising if the real world is quantum mechanical. And if the experimental
values agree, as seems to be the case, with quantum mechanics, what this would
seem to tell us is that our world is quantum mechanical and correctly described
using properties associated with a Hilbert space, rather than by classical
hidden variables.

\xb
\outl{If realism means: measurements reveal properties, then HSQM is realistic}
\xa

\xb
\outl{Physicist's belief that  `expt outcome $\lra$ previous micro property'
  supported by HSQM}
\xa

\xb
\outl{Describing micro properties using HV contradicts expts}
\xa

If we understand \emph{realism} to mean, in line with the commentary in
\cite{Wsmn15}, that measurements reveal pre-existing physical properties of
the world, Hilbert space quantum mechanics is realistic, as long as one does
not make the unreasonable demand that measurements reveal things which do not
exist in the real (quantum) world. Many physicists believe that the macroscopic
outcomes of their experiments indicate previous microscopic states-of-affairs,
and Hilbert-space quantum mechanics supports this conviction, as long as those
microscopic situations can be described using Hilbert subspaces. It is the
attempt to describe these pre-existing physical properties using classical
hidden variables, while ignoring noncommutativity, that disagrees with
experiment. The reasonable conclusion would seem to be that the real world is
quantum mechanical, not classical. Thus the term ``local realism'' in
\cite{Hnao15,Wsmn15} would be less misleading were it replaced by
``local \emph{classical} realism.''

\xb \outl{Is HSQM 'local' meaning NO `spooky action at a distance' (SPAAD)?} \xa

\xb \outl{SPAAD nonsignaling $\Ra$ no direct detection. Evidence comes from
  (possibly incorrect) theory} \xa

But is Hilbert-space quantum mechanics, the kind that agrees with experiment, a
\emph{local} theory? In answering this question we must be careful about what
we mean by ``local.'' Sometimes events that occur at widely separated places
can be correlated with each other in a way that indicates they have a common
cause in the past. This kind of nonlocality, common in classical physics, is
not surprising. What is of interest for the present discussion is whether
quantum theory allows for, or implies the existence of, \emph{nonlocal
  dynamical influences} whereby a cause at one location can produce an effect
at a different location when the two are at spacelike separation, i.e., the
events are sufficiently far apart that no signals can pass from one to the
other without exceeding the speed of light. Such influences are what Einstein
referred to as ``spooky action at a distance.''
Even those who believe, on the basis of violations of Bell inequalities, that
such superluminal influences exist will concede that they are
``non-signaling'': they cannot be used to convey information from one location
to another. This precludes any direct experimental test for their existence.
Instead the evidence is indirect: there are correlations that cannot be
explained by one's local theory, and from this one infers that the world is
nonlocal. But perhaps the problem lies with the theory.

\xb
\outl{Bell: local HV theory contrary to QT. He lacked good formulation of HSQM}
\xa

\xb
\outl{GMH, Hartle, Omnes, RBG $\ra$ CH; was unknown to Bell when he died}
\xa

\xb
\outl{CH probs compatible w Hilbert sspaces, imply Einstein
  locality $\Ra$ HSQM is local}
\xa

Bell argued that the correlations predicted by quantum mechanics, and confirmed
by experiment, cannot be explained by a local hidden variables theory. Alas, he
did not possess a consistent formulation of Hilbert-space quantum mechanics
that would have allowed him to compare its predictions with those provided by
local hidden variables. Standing in his way was the infamous \emph{measurement
  problem} of quantum foundations, which is to find a consistent and fully
quantum mechanical description of the physical processes that go on in a
measurement, starting with the microscopic property that is to be measured, and
ending with the macroscopic property (often called a ``pointer position'')
correlated with, and thus indicating, the earlier microscopic property. Bell
never solved this problem, as is evident from one of his last papers
\cite{Bll90}. Taken together with his other publications it shows that he was
unaware of, or at least had not given serious thought to, an approach developed
in the 1980s by Gell-Mann, Hartle, Omn\`es, and the undersigned, and which did
not reach a fully consistent form until the mid 1990s after Bell's death. It is
known as the \emph{consistent} or \emph{decoherent histories} formulation of
quantum mechanics, and it assigns probabilities in a way consistent with the
use of Hilbert subspaces, taking proper account of the noncommutativity of
quantum operators%
\footnote{The reader interested in these ideas will find presentations of
  moderate length in \cite{Hrtl11} and \cite{Grff14b}; an extended treatment is
  in \cite{Grff02c}}. %
An analysis based on this approach, with details given in \cite{Grff11}, shows
that quantum dynamics is consistent with the following principle of
\emph{Einstein locality}:
\begin{quote}
Objective properties of isolated individual systems do not change when 
something is done to another non-interacting system.
\end{quote}
This agrees with the characterization of ``local'' found in \cite{Wsmn15},
and shows that Hilbert-space quantum mechanics is local as well as being
in good agreement with experiment.

\xb
\outl{Delft results agree w realistic, local HSQM. SPAAD 
  can't carry info because it doesn't exist}
\xa

In summary, the impressive experimental results reported in \cite{Hnao15} are
in complete accord with, and indeed confirm, the validity of Hilbert-space
quantum mechanics, which is both realistic and local when proper account is
taken of the fact that the quantum world, the real world we live in, differs in
important respects from the world of classical physics. It is the use of
classical hidden variables in derivations of Bell inequalities which leads to
results in disagreement with quantum theory. And there is a very simple
explanation of why spooky action at a distance is unable to convey information:
it does not exist.

\xb
\section*{Acknowledgments}
\xa

Correspondence with R. Hanson, P. Hohenberg and H. Wiseman is gratefully
acknowledged. Some of the research on which the above conclusions are based
received support from the National Science Foundation through Grant 1068331.

\end{document}